\documentclass[prb, twocolumn, amsmath,amssymb, reprint]{revtex4-2}

\usepackage{amsmath}
\usepackage[english]{babel}

\usepackage{graphicx}
\usepackage{dcolumn}
\usepackage{bm}

\usepackage{mathptmx}
\usepackage{color}

\newcommand{\ds}{\displaystyle}

\begin{document}

\preprint{AIP/123-QED}

\title{Photogalvanic phenomena in superconductors supporting intrinsic diode effect}
\author{S. V. Mironov}
\affiliation{Institute for Physics of Microstructures, Russian Academy of Sciences, 603950 Nizhny Novgorod, GSP-105, Russia}
\author{A. S. Mel'nikov}
\affiliation{Moscow Institute of Physics and Technology (National Research University), Dolgoprudnyi, Moscow region, 141701 Russia}
\affiliation{Institute for Physics of Microstructures, Russian Academy of Sciences, 603950 Nizhny Novgorod, GSP-105, Russia}
\author{A. I. Buzdin}
\affiliation{University Bordeaux, LOMA UMR-CNRS 5798, F-33405 Talence Cedex, France}
\affiliation{World-Class Research Center ``Digital biodesign and personalized healthcare'', Sechenov First Moscow State Medical University, Moscow 119991, Russia}

\date{\today}

\begin{abstract}
In this work we suggest a phenomenological theory of photogalvanic phenomena in superconducting materials and structures revealing the diode effect.
Starting from a generalized London model including the quadratic nonlinearity in the relation between the supercurrent and superfluid velocity we show that the electromagnetic wave  incident on the superconductor can generate a nontrivial superconducting phase difference between the ends of the sample. Being enclosed in a superconducting loop such phase battery should generate a dc supercurrent 
circulating in the loop. Increasing the electromagnetic wave intensity one can provoke the switching between the loop states with different vorticities.
\end{abstract}

\maketitle

During the past three years, a lot of attention has been attracted to the superconducting diode effect (SDE) \cite{Nadeem_rev}. The nonreciprocal current--voltage (I--V) characteristics of diodes permit rectification of alternating currents which is very important for signal processing. The observation of diode effect in asymmetric superconducting devices was reported 25 years ago in spatially nonuniform Josephson junctions \cite{Krasnov} and 20 years ago in superconducting ratchets \cite{Villegas}. Various scenarios of broken inversion and time reversal symmetry in superconducting hybrid structures have been suggested in further works including asymmetry in vortex potentials, surface barriers, etc. \cite{silaev,lyu,majer,vodolaz,silva,suri}.
Interestingly, the SDE has been recently observed in bulk material-Nb/V/Ta superlattices without inversion center and in the presence of the in-plane magnetic field \cite{Ando}. A theoretical explanation of this SDE \cite{Daido, He} is based on the interplay between Rashba spin-orbit coupling (SOC) and Zeeman splitting which is responsible for the appearance of the additional type of invariants (linear and cubic on the order parameter gradient) in Ginzburg-Landau (GL) functional describing such a superconducting material. It was predicted that the presence of the linear in the order parameter gradient invariants in GL functional \cite{Edelstein_1, Edelstein_2} should generate the spontaneous current in different types of the non-uniform superconducting systems \cite{Pershoguba, Mironov_1, Robinson, Olde_Olthof, Kopasov_1, Kopasov_2} (also see, e.g., Ref.~[\onlinecite{Samokhvalov}] for review). Different microscopic approaches accounting intrinsic diode effect
have been discussed, e.g., in  Refs.~[\onlinecite{Wakatsuki,yuan,berg,karabas,kokk}].
Nonreciprocal phenomena associated with magnetochiral anisotropy were also observed in symmetric Josephson junctions \cite{Baumgartner, Jeon, Pal} and the corresponding theoretical descriptions are reviewed in [\onlinecite{Kochan}]. Josephson diodes based on the SQUID setup have been discussed recently in [\onlinecite{fominov}].

One of the important manifestations of the nonreciprocal transport properties in nonsuperconducting materials is connected with their unusual nonlinear electromagnetic response which should include the even harmonics of the frequency of the incident electromagnetic wave. The related photogalvanic phenomena have been studied intensively for several decades both theoretically end experimentally
 (see, e.g., Refs.~[\onlinecite{belinicher,ivchenko,glazov,nagaosa}] for review). The hallmark of these effects is known to be the generation of a dc current by the electromagnetic radiation, i.e. the phenomenon of rectification.

The goal of this Letter is to suggest a general phenomenological description of the rectification phenomenon in nonreciprocal superconducting materials and propose a simple setup for experimental study of the rectified current by measuring the superconducting phase gain induced by the electromagnetic radiation.
Note that the rectification phenomenon discussed below is different both from the inverse Faraday effect in superconductors requiring a circular polarized radiation \cite{Mironov_2, Croitoru, Plastovets} and from the effect of the photon drag in superconducting condensates studied recently in [\onlinecite{mironov2024}].

To elucidate the mechanism beyond the rectification effect we consider a superconducting film of the thickness $d_s\sim\xi_0$ ($\xi_0$ is the superconducting coherence length) placed on top of the ferromagnetic insulator with the exchange field directed parallel to the plane of the film. In the presence of strong Rashba spin-orbit coupling and Zeeman splitting of the electron spin bands arising due to the exchange field of the underlying ferromagnet the density $f({\bf r})$ of the Ginzburg - Landau (GL) free energy of the superconductor $F=\int f({\bf r})d^3{\bf r}$ reads 
\begin{equation}\label{GL_FE_general}
\begin{array}{c}{\ds
f({\bf r})=a|\psi|^2+\frac{\hbar^2}{4m}|\hat{\bf D}\psi|^2+\frac{b}{2}|\psi|^4+\frac{({\rm rot}{\bf A})^2}{8\pi}}\\{}\\{ \ds ~~~~~~~~~~ +\frac{\hbar^2}{4m} ({\bf n}\times {\bf h})\cdot[\psi^*\left(\varepsilon\hat{\bf D}+\eta\hat{\bf D}^3\right)\psi+{\rm h.c.}].}
\end{array}
\end{equation}
Here $a=-\alpha(T_c-T)$ and $b$ are the standard GL coefficients, $\psi$ is the superconducting order parameter, $\hat{\bf D}=-i\nabla+(2\pi/\Phi_0){\bf A}$ is the gauge-invariant momentum operator divided by the Planck constant $\hbar$, $\Phi_0=\pi\hbar c/e$ is the flux quantum (here $e>0$), ${\bf n}$ is the unit vector along the direction of broken inversion symmetry (for the superconducting film ${\bf n}$ is perpendicular to the film surface), ${\bf h}$ is the unit vector along the direction of the exchange field vector  responsible for the Zeeman-like spin splitting, and the constants $\varepsilon$ and $\eta$ describe the magnitude of different contributions associated with the Rashba SOC which are determined by both the SOC constant and the exchange field.
Taking the order parameter in the form $\psi= \rho e^{i\varphi}$  we find the relation 
\begin{equation}\label{Dpsi}
\hat{\bf D}\psi= \left(-i\nabla\rho+ \frac{2m}{\hbar}{\bf v}_s \rho \right)e^{i\varphi} ,
\end{equation}
where we introduced the notation 
\begin{equation}\label{vs_def}
{\bf v}_s=\frac{\hbar}{2m}\left(\nabla\varphi+\frac{2\pi}{\Phi_0}{\bf A}\right)
\end{equation}
 for the superfluid velocity.
Then the above free energy (\ref{GL_FE_general}) can be rewritten as a functional of the superfluid velocity using the perturbative expansion 
 of the order parameter absolute value $\rho$ in ${\bf v}_s$:
 \begin{equation}
 \label{GL_FE_general1}
f({\bf v}_s)=f_0\left({\bf v}_{s0}\right)+ m\rho_0^2\tilde{\bf v}^2_s
-m\rho_0^2\frac{({\bf u}_0\cdot\tilde{\bf v}_s)}{v_c}\tilde{\bf v}^2_s+\frac{({\rm rot}{\bf A})^2}{8\pi}.
\end{equation}
In Eq.~(\ref{GL_FE_general1}) ${\bf v}_{s0}\simeq -{\bf u}_0\hbar\varepsilon/2m$ is the superconducting velocity corresponding to the free energy minimum, $\tilde{\bf v}_s={\bf v}_s-{\bf v}_{s0}$ is the deviation of the supervelocity from ${\bf v}_{s0}$ in the presence of the superconducting current, 
the density $\rho_0^2$ takes the approximate value $\rho_0^2\simeq|a|/b$, the vector ${\bf u}_0= ({\bf n}\times {\bf h})$, and the constant $v_c$ is defined as $v_c=\hbar/(4m\eta)$. Note that since the expansion over $\tilde{\bf v}_s$ is performed in the vicinity of the free energy minimum, the linear in $\tilde{\bf v}_s$ term in the above expansion is absent. Note also that in Eq.~(\ref{GL_FE_general1}) we keep only the terms up to $O\left(\tilde{v}_s^3\right)$, $O\left(\varepsilon\right)$ and $O\left(\eta\right)$ while all the higher order terms are omitted.
Then the expression for the supercurrent density corresponding to the above functional reads:
 \begin{equation}\label{Curr_gen}
{\bf j}_s=-\frac{e}{m}\frac{\partial f}{\partial \tilde{\bf v}_s}=-e\rho_0^2\left[2\tilde{\bf v}_s-\frac{\tilde v_s^2}{v_c}{\bf u}_0 -2\frac{({\bf u}_0\cdot\tilde{\bf v}_s)}{v_c}\tilde{\bf v}_s \right] \ .
\end{equation}
 
 \begin{figure}[t!]
\begin{center}
\includegraphics[width=1\linewidth]{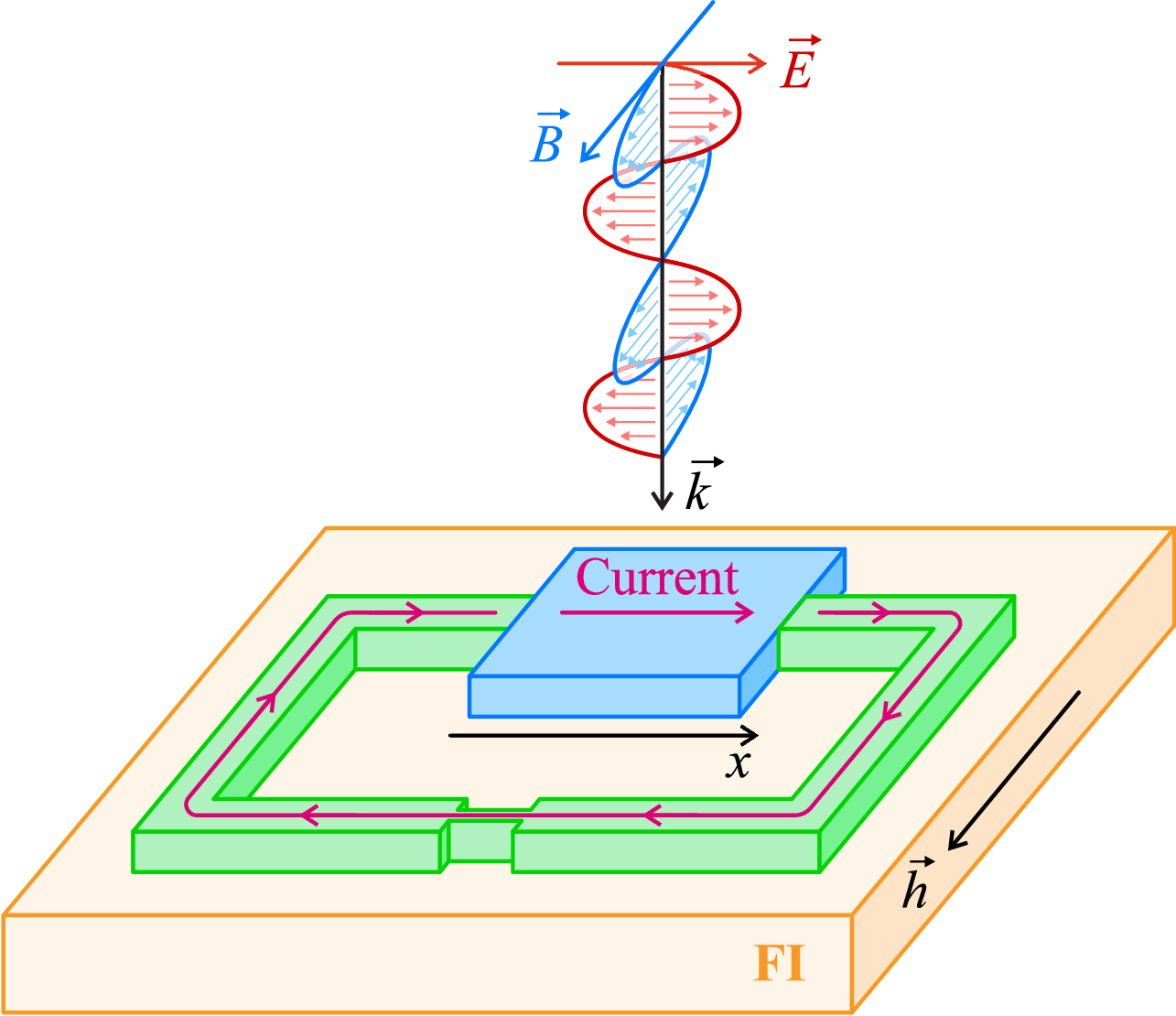}
\end{center}
\caption{Schematic setup for measurement of the rectified dc current induced by electromagnetic wave. The radiated superconducting sample with Rashba spin-orbit coupling is embedded into the superconducting loop. All parts of the loop are placed on top of the ferromagnetic insulator (FI) with the in-plane exchange field. The constriction of the superconductor plays the role of weak link.}\label{Fig1}
\end{figure}

Let us now apply this expression for the analysis of nonlinear electrodynamics and related nonreciprocal phenomena. For illustration of possible photogalvanic effects we consider a superconducting film of the thickness $d_s$ irradiated by the linearly polarized electromagnetic wave with the wave vector perpendicular to the film surface. The thickness $d_s$ is assumed to be much larger than the interatomic distance to ensure the full electromagnetic wave reflection but, at the same time, much smaller than the London penetration depth. The latter condition allows to neglect the spatial distribution of the optically-induced electric current across the film. For  further calculations it is convenient to consider the magnetic field of the incident wave in the plane of the film in the form ${\bf B}={\rm Re}\left({\bf B}_\omega e^{-i\omega t}\right)$ where ${\bf B}_\omega$ is the complex amplitude of the wave and $\omega$ is the wave frequency. Then integrating the Maxwell equation for ${\rm curl}~{\bf B}$ over the film thickness we get:
\begin{equation}\label{Curr_omega}
2\left({\bf n}\times {\bf B_\omega}\right)=\frac{4\pi}{c}{\bf j}_{s\omega}d_s,
\end{equation}
where ${\bf j}_{s\omega}$ is the complex amplitude of the supercurrent at the frequency $\omega$ and the factor $2$ in the l.h.s. accounts the doubling of the amplitude of the magnetic field at the sample boundary due to the full reflection of the incident wave. 
The total superfluid velocity can be written as an expansion in the harmonics at different frequencies:
\begin{equation}
 \tilde{\bf v}_s=\tilde{\bf v}_{0}+{\rm Re}\sum_{k=1}^\infty\left(\tilde{\bf v}_{k\omega}e^{-ik\omega t}\right).
\end{equation}
According to Eq.~(\ref{Curr_omega}), the velocity component with $k=1$ directly induced by the incident wave reads 
\begin{equation}\label{vs_omega}
\tilde{\bf v}_{\omega}=-\frac{c}{4\pi e\rho_0^2 d_s}\left({\bf n}\times {\bf B_\omega}\right),
\end{equation}
while the components $\tilde{\bf v}_{k\omega}$ with $k\neq 1$ are induced due to the SOC and have higher order in small parameters $\varepsilon$ and $\eta$. Within the perturbation approach one can write the following expression for the dc component of the supercurrent:
 \begin{equation}\label{js0}
{\bf j}_{s0}=-2e\rho_0^2\tilde{\bf v}_0+\frac{
e\rho_0^2}{2v_c}\left[|\tilde {\bf v}_\omega|^2{\bf u}_0 +({\bf u}_0\cdot\tilde{\bf v}^*_\omega)\tilde{\bf v}_\omega 
+({\bf u}_0\cdot\tilde{\bf v}_\omega)\tilde{\bf v}^*_\omega \right].
\end{equation}

The photo-induced dc current is described by the terms in square brackets and gives the source for the
dc superfluid velocity $\tilde{\bf v}_0$ depending on the appropriate boundary conditions for the full dc current. We assume here this source to be small and, thus, omit the quadratic in $\tilde{\bf v}_0$ contribution to the above supercurrent expression. Further analysis of the expression (\ref{js0}) depends on the polarization of the incident wave. Considering, first, the linearly polarized radiation we may put ${\bf B_\omega} = {\bf b} B_\omega$, where 
 ${\bf b}$ is a real valued unit vector. Then introducing the unit vector ${\bf e}={\bf n}\times {\bf b}$ which enters Eq.~(\ref{vs_omega}) and is directed along the electric field in the incident wave we may rewrite Eq.~(\ref{js0}) as
 \begin{equation}\label{js0_2}
{\bf j}_{s0}=-2e\rho_0^2\tilde{\bf v}_0+\frac{c^2\left|B_{\omega}\right|^2}{32\pi^2 e\rho_0^2 d_s^2v_c}\left[{\bf u}_0 +2({\bf u}_0\cdot{\bf e})~{\bf e} 
\right].
\end{equation}
Then it is convenient to rewrite $({\bf u}_0\cdot{\bf e}){\bf e}={\bf u}_0+{\bf e}\times\left({\bf e}\times {\bf u}_0\right)$ and take into account that ${\bf e}\times {\bf u}_0=-\left({\bf u}_0\cdot {\bf b}\right){\bf n}$ and ${\bf e}\times {\bf n}={\bf b}$ which finally gives 
 \begin{equation}\label{js0_3}
{\bf j}_{s0}=-2e\rho_0^2\tilde{\bf v}_0+\frac{c^2\left|B_{\omega}\right|^2}{32\pi^2 e\rho_0^2 d_s^2v_c}\left(3{\bf u}_0 -2\sin\theta{\bf b} 
\right),
\end{equation}
where $\sin\theta={\bf u}_0\cdot {\bf b}={\bf b}\cdot\left({\bf n}\times{\bf h}\right)$ with $\theta$ being the angle between the unit vectors ${\bf b}$ and ${\bf h}$ characterizing the directions of the magnetic field of the incident wave and the exchange field responsible for the Zeeman splitting of the electron energy bands, respectively. Interestingly, the angle between the vector $\left(3{\bf u}_0 -2\sin\theta{\bf b}\right)$ in Eq.~(\ref{js0_3}) and the vector ${\bf u}_0$ cannot exceed $\pi/6$ (this maximal value corresponds to $\theta=\pi/3$).

Provided the boundary conditions in the sample forbid the current flow we should put ${\bf j}_{s0}=0$. Integrating Eq.~(\ref{js0_3}) 
along a certain direction $x$ under the assumption that the sample size is much smaller than the London penetration depth (which allows us to neglect the Meissner screening of the current induced by the incident wave) we find that $\tilde{\bf v}_0=(\hbar/2m)\nabla\varphi$ and get a nonzero superconducting phase difference at the sample ends in the direction $x$:
\begin{equation}\label{phi_l}
\delta\varphi_x=\frac{8\pi^2\eta L_x  \left|B_\omega\right|^2}{B_\lambda^2 d_s^2 }
\left(3u_{0x} -2\sin\theta b_x\right),
\end{equation}
where $L_x$ is the irradiated sample size in the $x$ direction, the field $B_\lambda=\Phi_0/\lambda^2$ is of the order of the first critical field in superconducting material and 
$\lambda$ is the London penetration depth being equal to $\lambda=\sqrt{mc^2/(8\pi e^2 \rho_0^2)}$. Note that the full order parameter phase gain includes also the contribution $2m{v}_{s0}L_x/\hbar$ which appears even in the absence of the electromagnetic wave \cite{Robinson}.

\begin{figure}[t!]
\begin{center}
\includegraphics[width=0.95\linewidth]{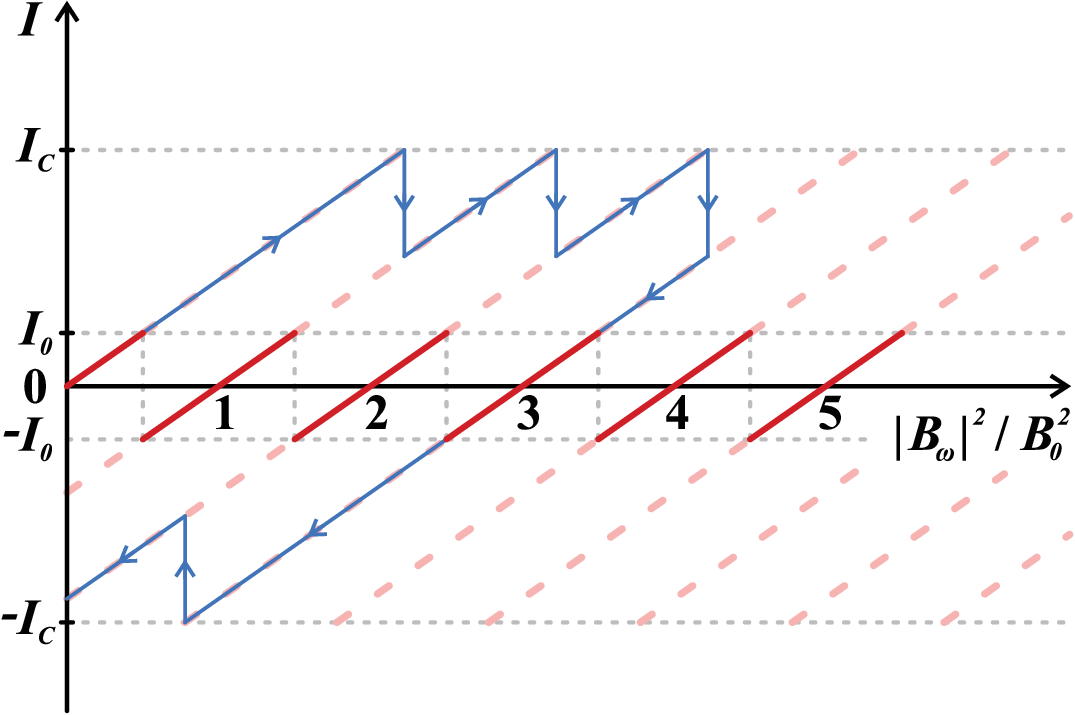}
\end{center}
\caption{Schematic plot illustrating the dependence of the rectified dc current on the intensity of the linearly polarized electromagnetic wave. Red solid lines correspond to the global energy minimum while the red dashed lines correspond to metastable states. The blue line shows an example of the hysteretic switching between the states induced by the sequential increase and decrease of the field intensity. Here $I_0=c\Phi_0/[2(L+L_k)]$, $I_c$ is the critical depairing current, and $B_0^2=
B_\lambda^2 d_s^2/\left[4\pi\eta L_x \left(3u_{0x} -2\sin\theta b_x\right)\right]$.}\label{Fig2}
\end{figure}

The described phenomenon takes place also for the incident wave of elliptical polarization. In this case the magnetic field in the wave can be written as ${\bf B}={\rm Re}\left[\left(B_\parallel{\bf w}_\parallel+iB_\perp{\bf w}_\perp\right)e^{-i\omega t}\right]$, where ${\bf w}_\parallel$ is the real valued unit vector along the semi-major axis in the plane of the superconducting film, ${\bf w}_\perp=\left({\bf n}\times{\bf w}_\parallel\right)$ is the unit vector perpendicular to ${\bf w}_\parallel$, while $B_\parallel$ and $B_\perp$ are the real amplitudes of the magnetic field in the direction of the semi-major and semi-minor axes. Then performing the analysis similar to the one for the linear polarization and taking into account that $\left[{\bf n}\times \left({\bf n}\times{\bf w}_\parallel\right)\right]=-{\bf w}_\parallel$ we find that the component $\tilde{\bf v}_\omega$ of the superconducting velocity oscillating at the frequency $\omega$ reads
\begin{equation}\label{vs_omega_2}
\tilde{\bf v}_{\omega}=-\frac{c}{4\pi e\rho_0^2 d_s}\left(B_\parallel{\bf w}_\perp-iB_\perp{\bf w}_\parallel\right),
\end{equation}
and the expression (\ref{js0}) for the dc component of the superconducting current after algebraic simplifications takes the form
\begin{equation}\label{js0_ell2}
\begin{array}{c}{\ds
{\bf j}_{s0}=-2e\rho_0^2\tilde{\bf v}_0+\frac{c^2}{32\pi^2 e\rho_0^2 d_s^2v_c}\left\{2\left(B_\parallel^2+B_\perp^2\right){\bf u}_0 +\right.}\\{}\\{\ds \left.+\left(B_\parallel^2-B_\perp^2\right)\left[\left({\bf h}\cdot{\bf w}_\perp\right){\bf w}_\parallel+\left({\bf h}\cdot{\bf w}_\parallel\right){\bf w}_\perp\right]\right\}.}\end{array}
\end{equation}
From this expression one sees that for the circularly polarized light with $B_\parallel=B_\perp$ the current always flows along the vector ${\bf u}_0$ while for the elliptical polarization the current direction does not coincides with the direction of ${\bf u}_0$. Similarly to the case of the linear polarization it is possible to integrate Eq.~(\ref{js0_ell2}) over the length of the sample and obtain the phase difference between the sample edges.

Note that the predicted phenomenon should take place both in clean and dirty superconductors. Indeed, the effect of disorder in the Rashba superconductor was studied in Ref.~\cite{Houzet}, where it was demonstrated that moderate disorder, with an electron mean free path $l\gtrsim\xi_0$, only weakly influences its parameters. Conversely, in cases of strong disorder $l\ll\xi_0$, the superconducting length is renormalized in a standard way $\xi_0\to\sqrt{\xi_0l}$, and anomalous terms in the Ginzburg-Landau functional (odd-order derivatives) decrease by a factor of $\sqrt{l/\xi_0}$. Recent theoretical studies on the superconducting diode effect in Rashba superconductors have also demonstrated its robustness against disorder \cite{berg}.

Note also that in our calculations we assume the intensity of the incident wave to be uniform along the irradiated part of the sample surface. At the same time, the possible inhomogeneity of the wave profile should be at a scale larger than the wavelength, which well exceeds the superconducting coherence length. Therefore, the resulting phase accumulation is determined by the integral of the local optically induced current along the irradiated superconductor.

Experimentally it is more convenient to detect  the above phase difference inserting the sample into a closed superconducting 
loop (somewhat similar geometry was considered previously, e.g., in Ref.~\onlinecite{Aronov}). The corresponding sample geometry is sketched in Fig.~\ref{Fig1}. We assume that the loop contains a weak link, e.g., a constriction or the insert of normal metal, so that the critical current of the loop (the maximal current which can flow through the loop without dissipation) is determined by the critical current associated with this weak link. Integrating the expression (\ref{js0}) along the loop and over the superconductor cross-section $\sigma$ we find:
\begin{equation}
I\ell =\frac{\hbar c^2\sigma}{8\pi e\lambda^2}  
(\delta\varphi_x +\delta\varphi_{s0} - 2\pi N - 2\pi \Phi/\Phi_0)
\ ,
\end{equation}
where $\ell$ is the loop length, $\Phi = LI/c$ is the magnetic flux through the loop, $L$ is the geometric inductance, and
the phase gain 
\begin{equation}
\delta\varphi_{s0}=\oint {\bf v}_{s0} d{\bf \ell}= -\frac{\hbar\varepsilon}{2m}\oint {\bf u}_0d{\bf \ell}
\end{equation}
depends on the texture of the exchange field along the loop. Note that we assume here $\sigma\ll\lambda^2$. Note that only the part of the loop made of the superconducting material with the intrinsic diode effect should contribute to the integral for the phase $\delta\varphi_{s0}$ and, thus, this phase depends on the 
size of the loop segment covered by the ferromagnet \cite{Robinson}. Covering the full loop by the ferromagnet with
homogeneous exchange field we get a uniform field ${\bf u}_0$ and the phase $\delta\varphi_{s0}$ vanishes.
Introducing the kinetic inductance of the loop $L_k=4\pi\lambda^2\ell/\sigma$ we get
\begin{equation}\label{philoop}
I =\frac{c\Phi_0}{L+L_k}  \left(\frac{\delta\varphi_x+\delta\varphi_{s0} }{2\pi}-N\right).
\end{equation}
The direction of the $x$ axis is chosen along the tangential direction to the loop in the irradiated segment.
The integer number $N$ denotes the number of vortices entering the loop one by one with the increase in the phase gain $\delta\varphi_x$.
This vorticity number can be determined if we consider the magnetic energy of the loop
\begin{equation}
E_N = \frac{(L+L_k)I^2}{2c^2}=\frac{\Phi_0^2}{2(L+L_k)^2} \left(\frac{\delta\varphi_x}{2\pi}-N\right)^2.
\end{equation}
The condition $E_N = E_{N+1}$ gives us the phase gain values $\delta\varphi_x = 2\pi (N+1/2)$ defining the
electromagnetic wave amplitude $B_{\omega, N+1}$ corresponding to the switching between different number of vortices in the loop. 
In Fig.~\ref{Fig2} we show schematically the resulting behavior of the dc
current induced in the loop vs the electromagnetic wave intensity. Note that these transitions between different vortex states are in principle hysteretic so that experimentally 
the wave amplitudes giving the vortex entry and exit can be different (in Fig.~\ref{Fig2} we schematically show an example of such hysteretic behavior with blue lines). 
The estimate of the minimal intensity required to induce the first vortex state can be given as follows:
\begin{equation}
\left|B_\omega\right|^2=\frac{B_\lambda^2 d_s^2 (L+L_k)I_c}{4\pi\eta L_x c\Phi_0\left(3u_{0x} -2\sin\theta b_x\right)},
\end{equation}
where the critical current $I_c$ should be small enough to ensure the applicability of the perturbative approach used to derive Eq.~(\ref{js0}). Such generation of the vortex state is reversible: the decrease in $\left|B_\omega\right|^2$ should result in the decrease in the number of the trapped vortices. This peculiar effect provides a convenient tool for superconducting optofluxonics allowing one to manipulate the magnetic flux trapped inside the loop which is promising for applications in the devices of the rapid single flux quantum (RSFQ) logics.

Alternatively, the vortex entry into the loop can be associated with thermal fluctuations (see, e.g., \cite{Arutyunov}). In the absence of radiation the ensemble of identical loops should reveal the symmetry in the number of loops with positive and negative vorticities. At the same time, switching on the radiation should break this symmetry down.

To observe the predicted phenomena it is reasonable to use microwave radiation with the frequency which is small enough not to induce the oscillations of the normal charge carriers density. Since the induced phase difference $\delta\varphi_x\propto \left|B_\omega\right|^2$ the increase in the wave intensity should facilitate the generation of the  state with the nontrivial vorticity in the loop. To estimate the magnitude of the predicted effect we take into account that $\eta\sim\hbar v_{s0} E_{ex}/T_c^2\sim \left(E_{ex}/T_c\right)\left(v_{s0}/v_F\right)\xi_0$ \cite{Levchenko}, where $v_{s0}$ is the characteristic velocity describing the strength of spin-orbit coupling,  $E_{ex}$ is the exchange field induced in the superconductor by the underlying ferromagnetic insulator, $v_F$ is the Fermi velocity, and $\xi_0\sim \hbar v_F/T_c$ is the superconducting coherence length. For $E_{ex}/T_c\sim 10$, $v_{s0}/v_F\sim 0.1$ we find $\eta\sim \xi_0$. Then considering the radiation of the intensity $c\left|B_\omega\right|^2/8\pi\sim 10~ {\rm \mu W}/{\rm \mu m}^2$ as well as the superconductor with $d_s\sim\xi_0\sim 100~{\rm nm}$ and the critical magnetic field $B_\lambda\sim 10^{-2}~ {\rm T}$ from Eq.~(\ref{phi_l}) one gets the estimate $\partial \varphi/\partial x\sim 0.1~ {\rm \mu m}^{-1}$. At the same time, one should make sure that the radiation does not destroy the superconductivity in the loop due to the heating effects. To organize the effective heat removal from the sample one may use, e.g., the sapphire substrate. The estimates show \cite{Mironov_2} that at temperatures $T\sim 10~{\rm K}$ the sapphire substrate of the thickness $\sim 1~{\rm \mu m}$ with the temperature elevation of $\Delta T\sim 10~{\rm K}$ between its edges can support the heat transfer of the surface power
density up to $q\sim 10^3 {\rm \mu W}/{\rm  \mu m}^2$. Thus, the heating effects do not provide rigid restrictions for the observability of the described effects.

Note in conclusion that very similar experimental setup can be used to detect the phase gain induced by the photon drag effect 
discussed recently in [\onlinecite{mironov2024}]. The only difference will be in the form of the expression for the phase gain $\delta\varphi_x$
which will appear in this case only for the electromagnetic wave incident at superconductor film surface at a certain nonzero angle.

To sum up, we have suggested a phenomenological description of possible photogalvanic phenomena in superconducting systems with the intrinsic diode effect and analyzed an exemplary experimental setup which allows to observe the dc superconducting phase gain and supercurrent induced 
in these systems in the field of electromagnetic wave with different polarization.



\bigskip
\begin{acknowledgments}
This work was supported by the Russian Science Foundation (Grant No. 20-12-00053) in part related to the calculation of the photoinduced phase gain for different polarizations. The part of the work related to the analysis of photogalvanic current in the ring geometry was supported by Ministry of Science and Higher Education of the Russian Federation within the framework of state funding for the creation and development of World-Class Research Center (WCRC) ``Digital biodesign and personalized healthcare'' (Grant No. 075-15-2022-304), ANR SUPERFAST and the LIGHT S\&T Graduate Program. S. V. M. acknowledges the financial support of the Foundation for the Advancement of Theoretical Physics and Mathematics BASIS (Grant No. 23-1-2-32-1).
\end{acknowledgments}

\end{document}